\documentclass[aip,jcp,preprint,longbibliography]{revtex4-1}

\usepackage{color}
\usepackage{amsmath}
\usepackage{amssymb}
\usepackage{graphicx}
\usepackage[T1, OT1]{fontenc}
\usepackage{hyperref}
\usepackage{float}

\begin{document}

	\title{New quantum neural network designs}
	\author{Felix Petitzon	\footnote{ felix.petitzon@mail.com}}

	\begin{abstract}
		
		Quantum computers promise improving machine	learning. We investigated the performance of new quantum neural network designs. Quantum neural networks currently employed rely on a feature map to encode the input into a quantum state. This state is then evolved via a parameterized variational circuit. Finally, a measurement is performed and post-processed on a classical computer to extract the prediction of the quantum model. We develop a new technique, where we merge feature map and variational circuit into a single parameterized circuit and post-process the results using a classical neural network. On a variety of real and generated datasets, we show that the new, combined approach outperforms the separated feature map \& variational circuit method. We achieve lower loss, better accuracy, and faster convergence.

	\end{abstract}
	\maketitle
	
	\section{Introduction}
	On classical, conventional computers, neural networks (a class of supervised models) have been successfully employed in the last decade. Neural network models have achieved human or super-human performance on tasks such as image classification \cite{Dai2021} \cite{Yoo2015}, text translation \cite{deepl}, playing boardgames \cite{Silver2016}, and predicting how proteins folds \cite{Jumper2021}.

	\begin{figure}[H]
		\begin{center}
			\includegraphics[angle=0, width=1 \textwidth]{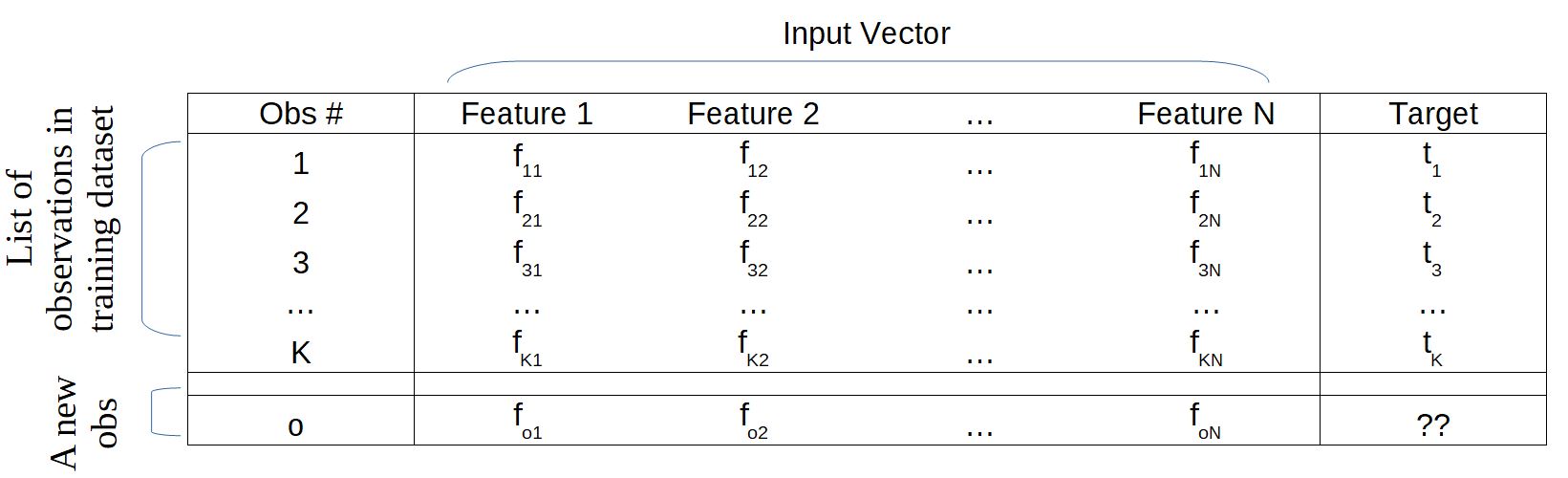}
			\caption{ Supervised learning.
				A training dataset consists of a list of observations. Each observation contains an input vector and a target. The task of a supervised model consists in learning from the training dataset how to predict the target given the input vector. The goal is to obtain a model that predicts accurately on novel inputs. That is, the model generalizes from the training dataset to new observations. }
		\end{center}
	\end{figure}	

	Quantum neural networks are the equivalent of these classical neural networks but are deployed on quantum computers. Quantum computers were first proposed in 1982 by Feynman \cite{Feynman1982} on the ground that exact simulations on classical computers grow exponentially in the space-time volume of a physical system. In contrast, those quantum computers could simulate nature requiring only proportional computational resources. In the mean time, quantum algorithms have been found with exponential speedups over their best known classical counterparts \textendash \hspace{0.03cm} notably prime number factorization \cite{shor96}; other quantum algorithms have quadratic speedups \cite{grover96}. These achievements, as well as the knowledge that classical computers cannot efficiently simulate quantum computers, have stirred the hope that quantum computers can lead to a speedup in quantum neural networks as well. 
	
	With the creation in recent years of the first physical quantum computers \cite{IBM2022} \cite{Rigetti2022} \cite{IonQ2020}, these quantum algorithms have been run on actual devices. However, these first quantum computers are noisy, intermediate-scale quantum computers (also called NISQs \cite{Preskill2018}). These limited devices only possess 10-100's of qubits, and thus only small quantum circuits can run on them. Furthermore, due to noise (i.e., errors when applying operations), these circuits need to be shallow, i.e. contain only limited operations. The advent of these NISQs has spurred the development of a new class of quantum algorithms that could realistically be run on such devices. These new algorithms are effective with a low number of qubits, have low circuit depths, and are resilient to errors. This encourages research into quantum neural networks that could already run on NISQs and provide speedups over classical counterparts. 
		
\begin{figure}[H]
		\begin{center}
			\includegraphics[angle=0,width=1 \textwidth]{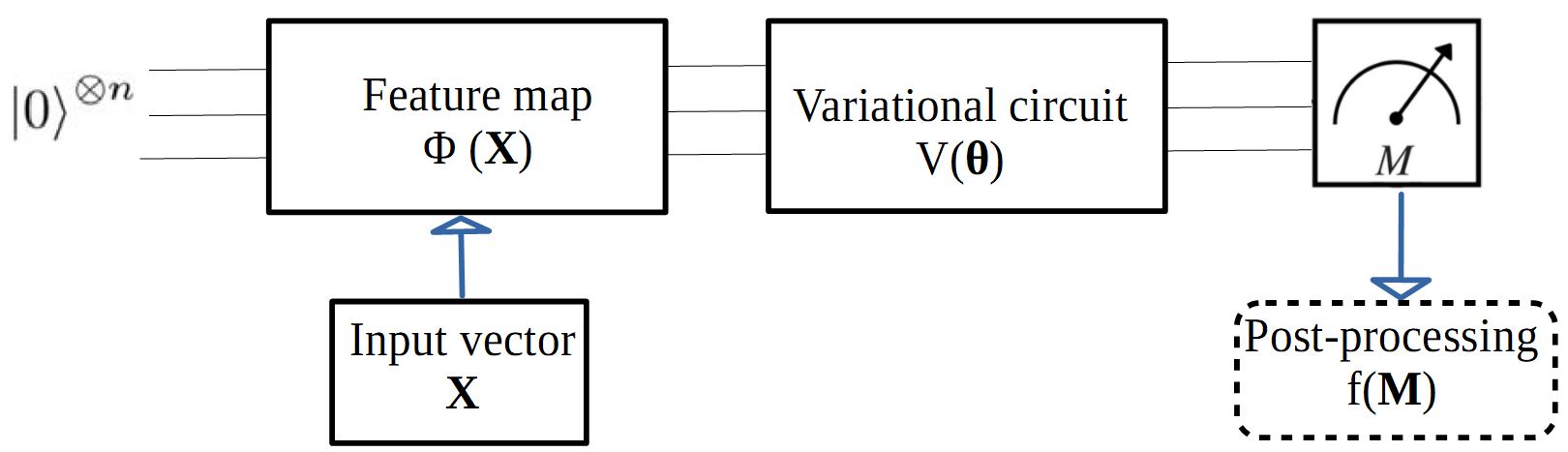}
			\caption{ Typical quantum neural net. For each observation, the input vector X is encoded into the statevector of a quantum circuit via a feature map $\Phi(X)$. This feature map carries out a set of rotations and entanglements which depend on the input fed to it. (Note: the input X could be pre-processed by a classical computer/neural net to reduce the number of features to a more manageable size.) Subsequently, a parameterized variational circuit $V(\theta)$ evolves the state, where $\theta$ is a list of parameters. Finally, the measurement M of the circuit is post-processed on a classical computer into an $output = f(M)$. This is the prediction of the quantum neural network. }
			\label{fig:qnn}
		\end{center}
	\end{figure}

	The proposed structure of such a quantum neural network is described in Fig \ref{fig:qnn} \cite{IBM_course_2022} \cite{Benedetti2019} \cite{Schuld2018}.
	
	The quantum neural network, similar to its classical analog, is trained on a dataset. A loss function is computed by comparing the predictions of the model with the target data. An optimizer \textendash \hspace{0.03cm} that runs on a classical computer \textendash \hspace{0.03cm} searches for the minimum loss by repeatedly running the quantum circuit and varying the parameters $\theta$. For a quantum neural network, it will optimize the parameters that define the rotations and entanglements of the quantum circuit. 	
	
	The choice of feature map and variational circuit is crucial in such design. Typically, the feature map is not parameterized and depends solely on the input. (Though, the idea to parametrize the feature map was envisioned before \cite{Lloyd2020}.) $\Phi(X)$ is rather based on some physical intuition or inspired by quantum chemistry \cite{Cerezo2021}. This puts a lot of burden on humans to find clever feature- and variational circuits.		
	
	\section{Approach}
	
	We propose that the split into a feature and variational circuit is an unnecessary and artificial division. This two-step approach seems to originate from the thinking that we need to first ``inject/encode the classical information" onto the quantum circuit, and then have a parameterized circuit provide ``a prediction on the encoded data." \cite{Havlicek2019}
	
	Neural networks on classical computers rely on few human inputs (merely a high-level structure of the model) and then crunch huge amounts of data. They rely heavily on statistics over analytical thinking; lots of empirical data over humans doing analytical work. In the same line of thought, we propose to merge the feature map and variational circuit into a single parameterized circuit, which depends on the input vector and a set of parameters $\theta$. We note here that the quantum device runs the entire circuit at once, and that separation into feature map and variational circuit is irrelevant to the compiler.
	
	 We also envision that the post-processing happens with a parameterizable classical model, instead of a fixed post-processing function. 
	 
	 \begin{figure}[H]
	 	\begin{center}
	 		\includegraphics[angle=0,width=1 \textwidth]{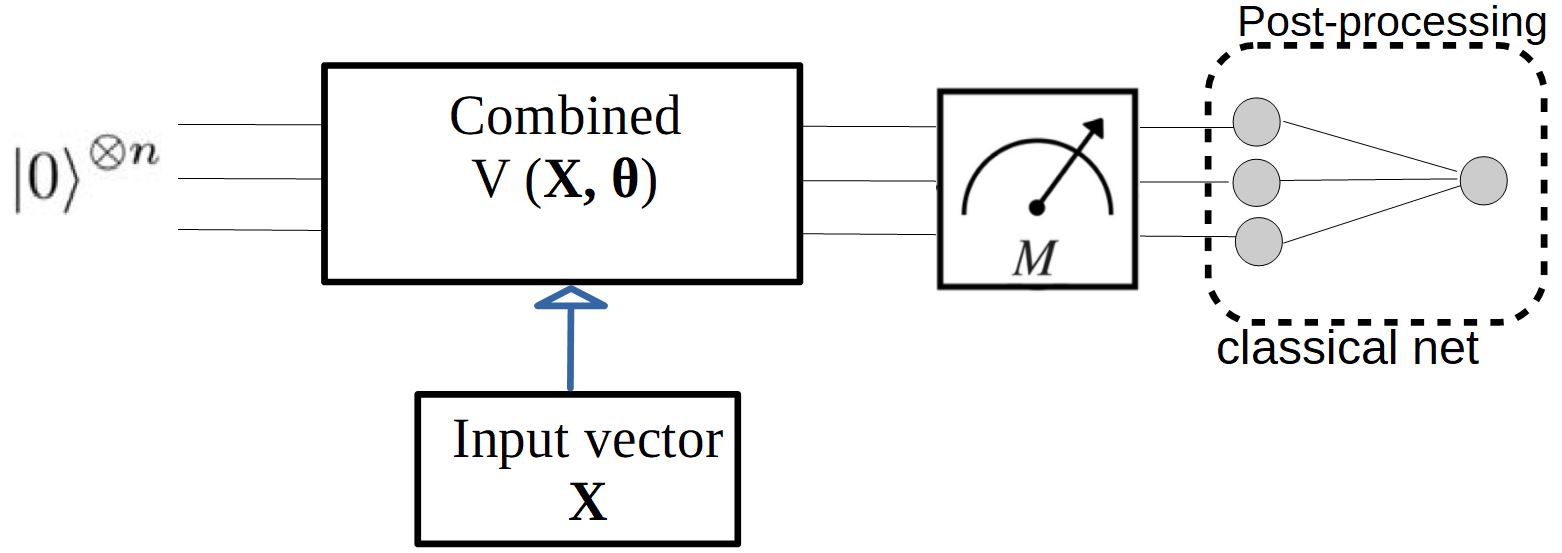}
	 		\caption{ New design of quantum neural net. For each observation, a parameterized circuit $V(X, \theta)$ is created, where X is the input vector. $\theta$ is a list of parameters which the optimizer will vary to search for the minimum loss. We call this the combined circuit. Next, each qubit is measured. Finally, these measurements M are post-processed by feeding them into a parametrizable classical neural network. This is the output of the quantum-classical neural net tandem.  }
	 		\label{fig:qnn_combined}
	 	\end{center}
	 \end{figure}	
	 
	 The first big difference compared to previously designed quantum neural nets is the combination of feature and variational parts into one. This effectively fully parametrizes the feature map and leaves it up to the optimizer to figure out the best combined circuit.
	 Secondly, we parametrize the post-processing part. Namely, instead of relying on a fixed function, we use a classical neural net sequentially after the quantum neural net. We dub this the quantum-classical tandem. We could go even further by using this as a building block and create an overall architecture that consists of repeating this block (quantum - classical) N times. Deep classical nets are able to approximate any function \cite{Gelenbe1999}. No such general proof, however, is known for quantum computers. By mixing a quantum net (=qnn) with a classical one (=cnn), we hope to bring out the strengths of both.

	This line of thought opens the door to new forms of qnns. High-level design by humans is still necessary to define the overall circuit. But the bulk of the work should be done by the different qnn and cnn `building blocks'. The idea of using multiple cnn and qnn blocks is approached in the existing TensorflowQuantum\cite{Broughton2020}, although the authors focus more on the possibilities of their software library and don't motivate the choice of mixing cnn and qnn blocks.
	
	We, in contrast, explicitly desire to mix cnn and qnn blocks. Our reasoning is: 1) the qnn blocks don't have known, proven speedups over the classical counterparts. 2) But qnn with entanglement gates cannot efficiently be simulated by classical computers, so some quantum speedup potential exists. 3) By blending both cnn and qnn, we intend the cnn to find out the correct way of feeding information to and extracting from the qnn. Hence we propose this hybrid quantum-classical neural network, which makes it possible to utilize existing quantum computers to their fullest extent.

	\section{Results }
	We test different models on a three datesets. The models are trained on a subset of the data. The loss and accuracy of each model are then assessed on a validation dataset, comprising observations not used in training. Only results for this validation dataset are reported. We test 5 different models: one classical, two quantum, and two combined classical/quantum. The quantum part is done via a simulation, and not directly run onto a NISQ device.
	
	\subsection{Models }
	
	\textbf{The classical neural net} \par
	The classical neural network, \textit{classical net}, that is used is the following:	
	\begin{figure}[H]
		\begin{center}
			\includegraphics[angle=0,width=0.67 \textwidth,origin=c  ]{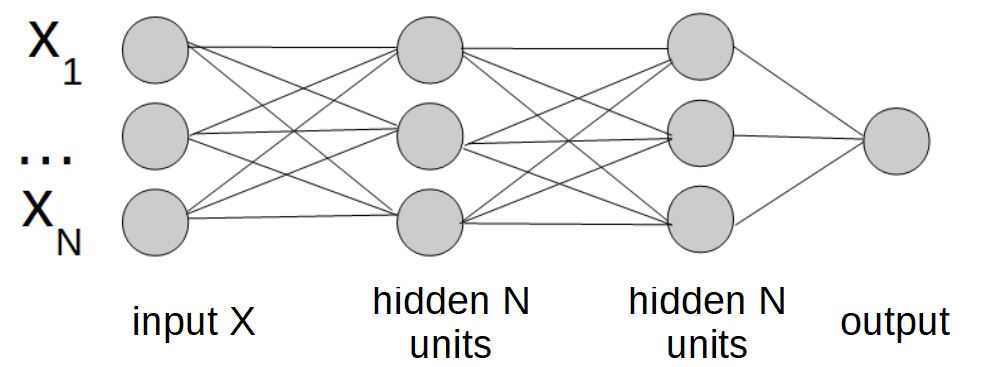}
			\caption{ Classical neural network. The input vector X of size N is first provided to a hidden layer. The output is then fed to a second hidden layer of equal size. Finally, a dense layer of size 1 produces the output. Omitted from this figure are any activation functions. We use leaky relu for regression, and tanh and the sigmoid activation function for classification problems. }
		\end{center}
	\end{figure}	
	\textbf{The feature and variational quantum neural net} \par
	
	The quantum neural network design presented in FIG. \ref{fig:qnn}. serves as a basis. More specifically, we borrow from an architecture proposed and motivated by Havlicek (2019) \cite{Havlicek2019}. The feature map is a ZzFeatureMap with 2 repetitions, as defined in IBM Qiskit \cite{ZzFeatureIbm}. The variational circuit is a RealAmplitudes circuit, as defined by IBM Qiskit \cite{RealAmpIbm}. We test using the parity of the measured qubits is used as post-processing function, and call this model \textit{featureVar}. We also test appending a classical net, which does the post-processing; we dub this \textit{featureVar + cnn}. \\
	
	\textbf{The combined quantum neural net} \par	
	
	A form of the quantum neural network design is presented in FIG. \ref{fig:qnn_combined}. Many different combined variational circuits $V(X, \theta)$ could be proposed. A multitude of rotation gates are applied, depending on either $X_{i}$, or some $\theta_{i}$, or a product of both. Entangling gates are applied as well.
	
	Some practical considerations need to be kept in mind, in particular hardware efficiency of NISQ devices. Notably: circuit depth needs to be limited, entangling gates generate the most errors, and the connectivity of the device needs to be considered. Nonetheless, the parameterized circuit must be expressive and entangling gates need to be included. Otherwise the quantum circuit could efficiently be simulated by a classical computer and hence provide no advantage.
	
	The exact design is provided in the Appendix. We call this circuit the \textit{combined qnn} when the parity of the measured qubits is used as output. When \textendash \hspace{0.03cm} instead of parity \textendash \hspace{0.03cm} a classical net is appended to the quantum circuit and delivers the final output, we name this model \textit{combined qnn + cnn}. This corresponds with the design presented in FIG. \ref{fig:qnn_combined}. \\

	\subsection{Results on Generated Classification Dataset}
	
	We generate a random dataset with the make\_classification function from the sklearn python library. The dataset contains 350 observations x 3 features and a (0,1) target. Out of the three, two are informative and one is a useless feature. This forms a binary classification problem. 
	
	\begin{figure}[H]
		\begin{center}
			\includegraphics[height=3.2in,width=3.2in,angle=0]{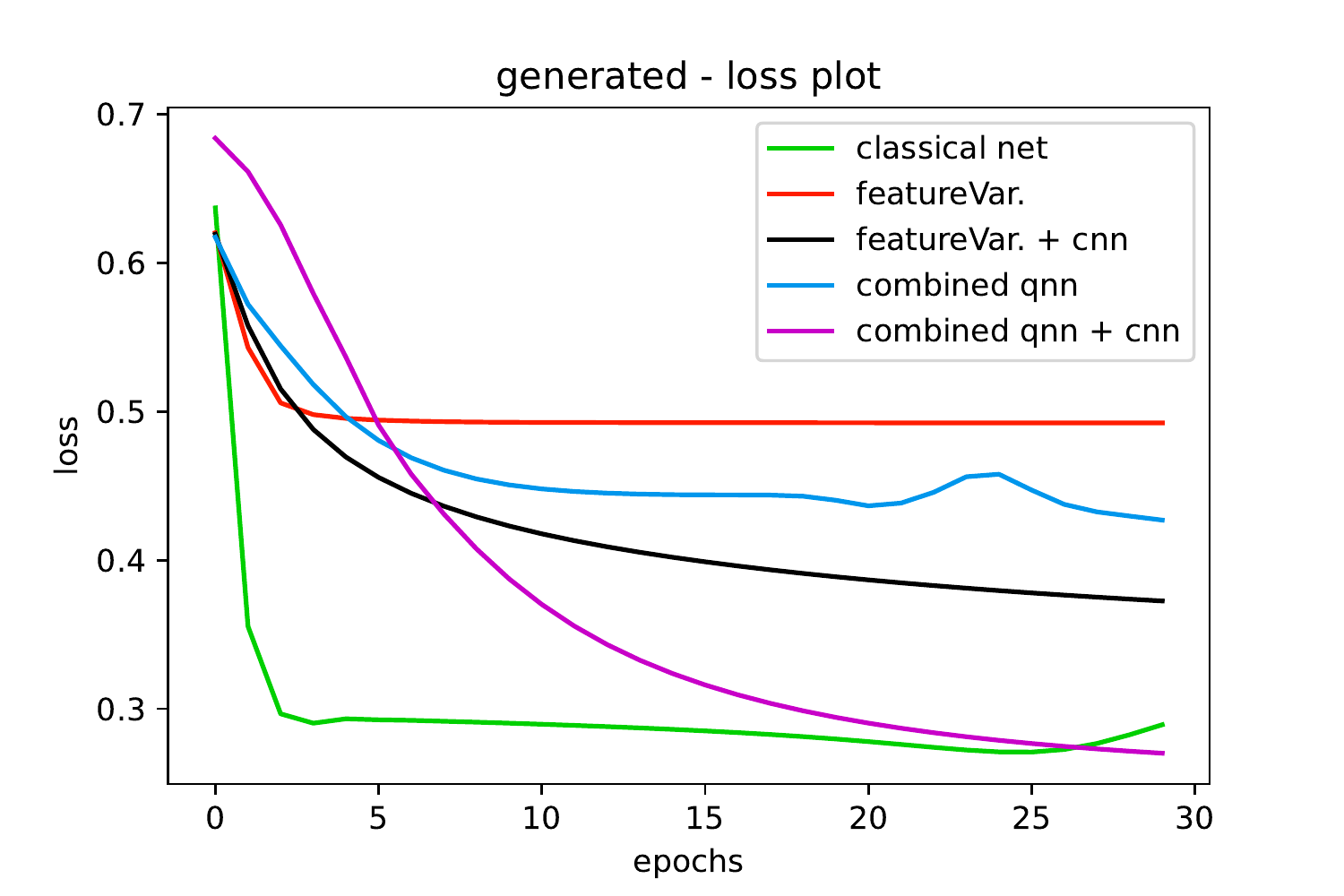}
			\includegraphics[height=3.2in,width=3.2in,angle=0]{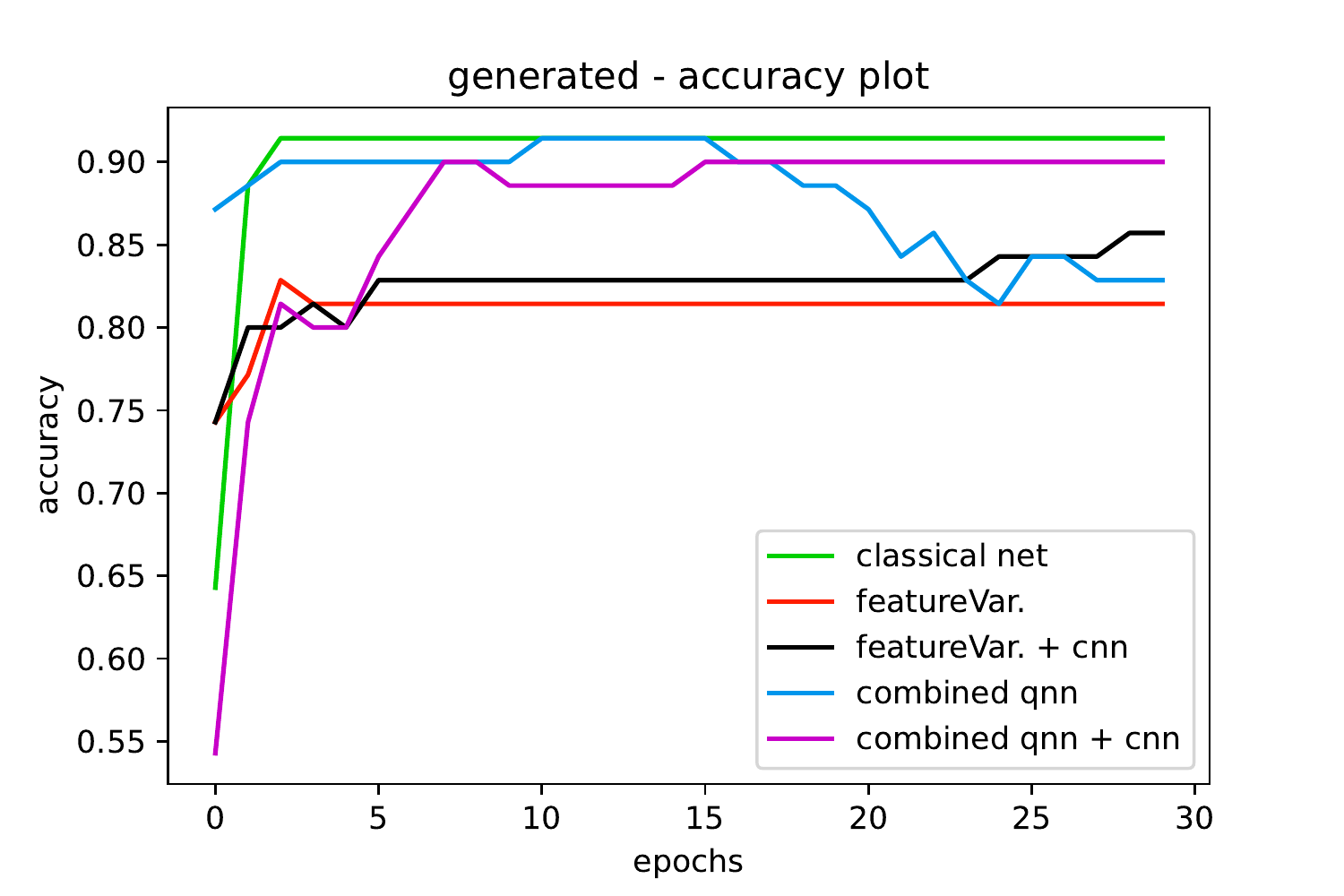}
			\caption{ Plot of the generated dataset results. The binary cross entropy loss (left) and accuracy (right) on the validation set.}
		\end{center}
	\end{figure}
	
	The \textit{classical net} achieves a loss of 0.271 on the validation set, and an accuracy of 91.4\%. In contrast, the \textit{featureVar.} model performs poorer. By adding a classical net as postprocess to this qnn (\textit{featureVar. + cnn}), the performance improves but remains below the \textit{classical net}. The \textit{combined qnn} achieves better accuracy and lower loss. And the \textit{combined qnn + cnn} has a performance about equal to the \textit{classical net}, despite less than half of the parameter count.
	
	\begin{table}[H]    
		\begin{tabular}{||c ||c | c | c||} 
			\hline
			model name & \# of classical / quantum parameters  & min loss & max accuracy \\ 
			\hline\hline
			classical net & 28 / 0 & 0.271 & 91.4\%  \\ 
			\hline
			featureVar. & 0 / 6 & 0.492 & 81.4\%  \\
			\hline
			featureVar. + cnn & 4 / 6 & 0.373 & 85.7\%  \\
			\hline
			combined qnn & 0 / 9 & 0.427 & 91.4\%  \\
			\hline
			combined qnn + cnn & 4 / 9 & 0.270 & 90.0\%  \\ [1ex] 
			\hline
		\end{tabular}
	\caption{\label{tab:genereated_ds} Results for the generated dataset. For each model, the number of classical and quantum parameters, the best loss and the best accuracy is shown.}
	\end{table}

	\subsection{Results on Diabetes Classification Dataset}
	
	This is a real-life dataset from a Kaggle competition \cite{diabetes2022}. This dataset is originally from the National Institute of Diabetes and Digestive and Kidney Diseases. The objective of the dataset is to diagnostically predict whether or not a patient has diabetes, based on certain diagnostic measurements included in the dataset. It contains 768 observations x 3 features and a (0,1) target. We only use a subset of features (3 out of 8) on this dataset for two reasons. Firstly, to speed up the computation time as the simulation of quantum circuits is otherwise slow. Secondly, using all the features of the dataset leads to `acing', i.e. reaching 95-99\% accuracy. The final score is then often determined by a few outlier points. And this makes comparing the relative performance of the models difficult. This is a binary classification problem.
	
	\begin{figure}[H]
		\begin{center}
			\includegraphics[height=3.2in,width=3.2in,angle=0]{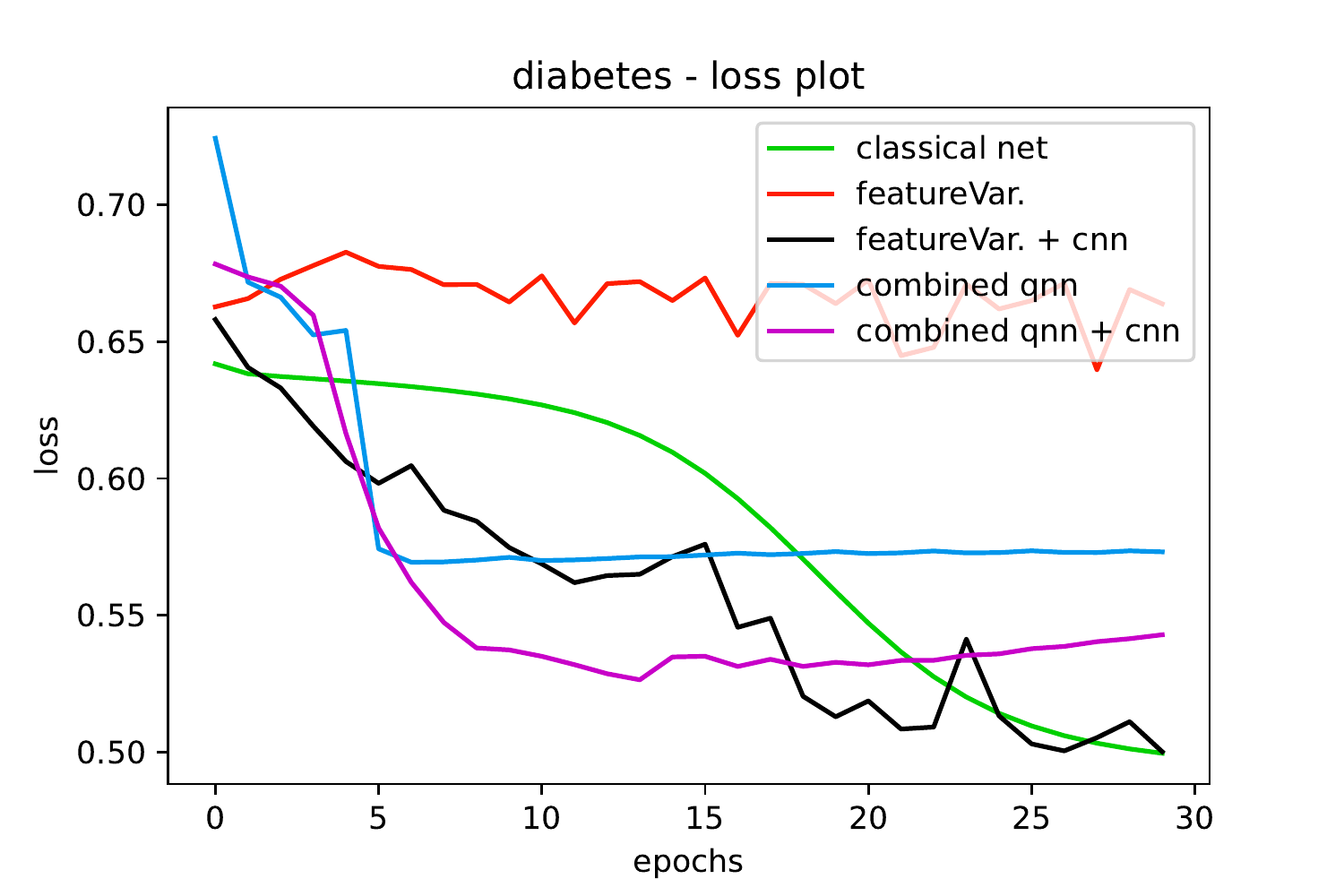}
			\includegraphics[height=3.2in,width=3.2in,angle=0]{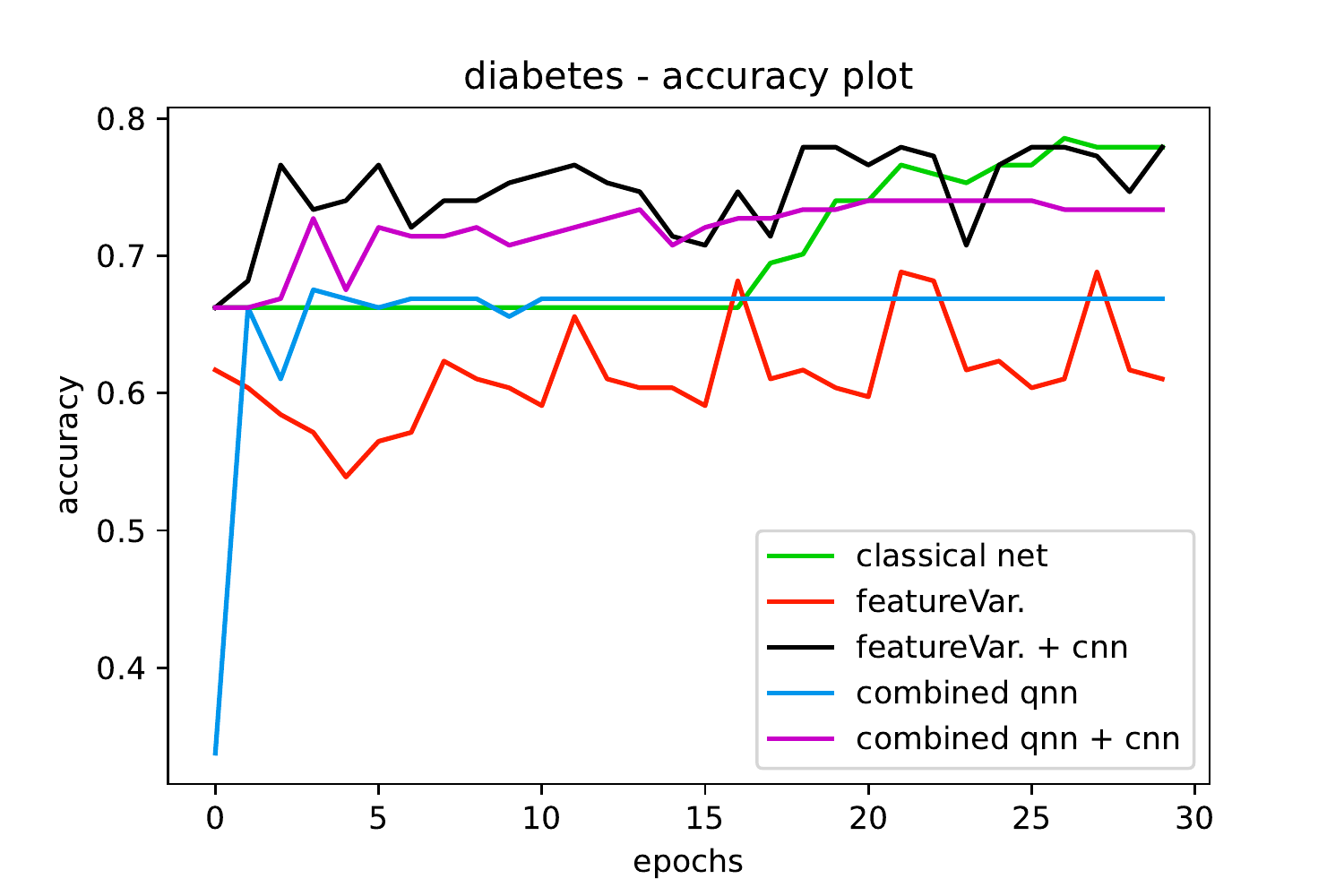}
			\caption{ Plot of the Diabetes dataset results. The binary cross entropy loss (left) and accuracy (right) on the validation set.}
		\end{center}
	\end{figure}

	The \textit{classical net} achieves a loss of 0.500 on the validation set, and an accuracy of 78\%.  The \textit{featureVar.} performs the poorest. By adding a classical net to postprocess this qnn (\textit{featureVar. + cnn}), the performance improves drastically and is on par with the \textit{classical net}, despite a lower parameter count. The \textit{combined qnn} and \textit{combined qnn + cnn} score below that.

	\begin{table}[H]
		\begin{tabular}{||c ||c | c | c ||} 
			\hline
			model name & \# of classical / quantum parameters  & min loss & max accuracy \\ 
			\hline\hline
			classical net & 28 / 0 & 0.500 & 77.9\%  \\ 
			\hline
			featureVar. & 0 / 6 & 0.640 & 68.8\%  \\
			\hline
			featureVar. + cnn & 4 / 6 & 0.500 & 78.0\%  \\
			\hline
			combined qnn & 0 / 9 & 0.569 & 66.9\%  \\
			\hline
			combined qnn + cnn & 4 / 9 & 0.529 & 74.0\%  \\ [1ex] 
			\hline
		\end{tabular}
	
	\caption{\label{tab:diabetes} Results for the Diabetes dataset. For each model, the number of classical and quantum parameters, the best loss and the best accuracy is shown. }
	\end{table}

	\subsection{Results on Banknote Fraud Classification Dataset }

	This is real-life dataset \cite{Dua:2019} based on images of genuine and forged banknotes. Wavelet Transform tool were used to extract features from the images. The dataset contains 1372 observations x 2 features and a binary (0,1) target. We only use a subset of features (2 out 4) on this dataset to speed up the computation time and to avoid `acing'. (i.e. reaching 95-99\% accuracy where the final score is then often determined by a few outlier points, which makes comparing the relative performance of the models difficult.)
	
	\begin{figure}[H]
		\begin{center}
			\includegraphics[height=3.2in,width=3.2in,angle=0]{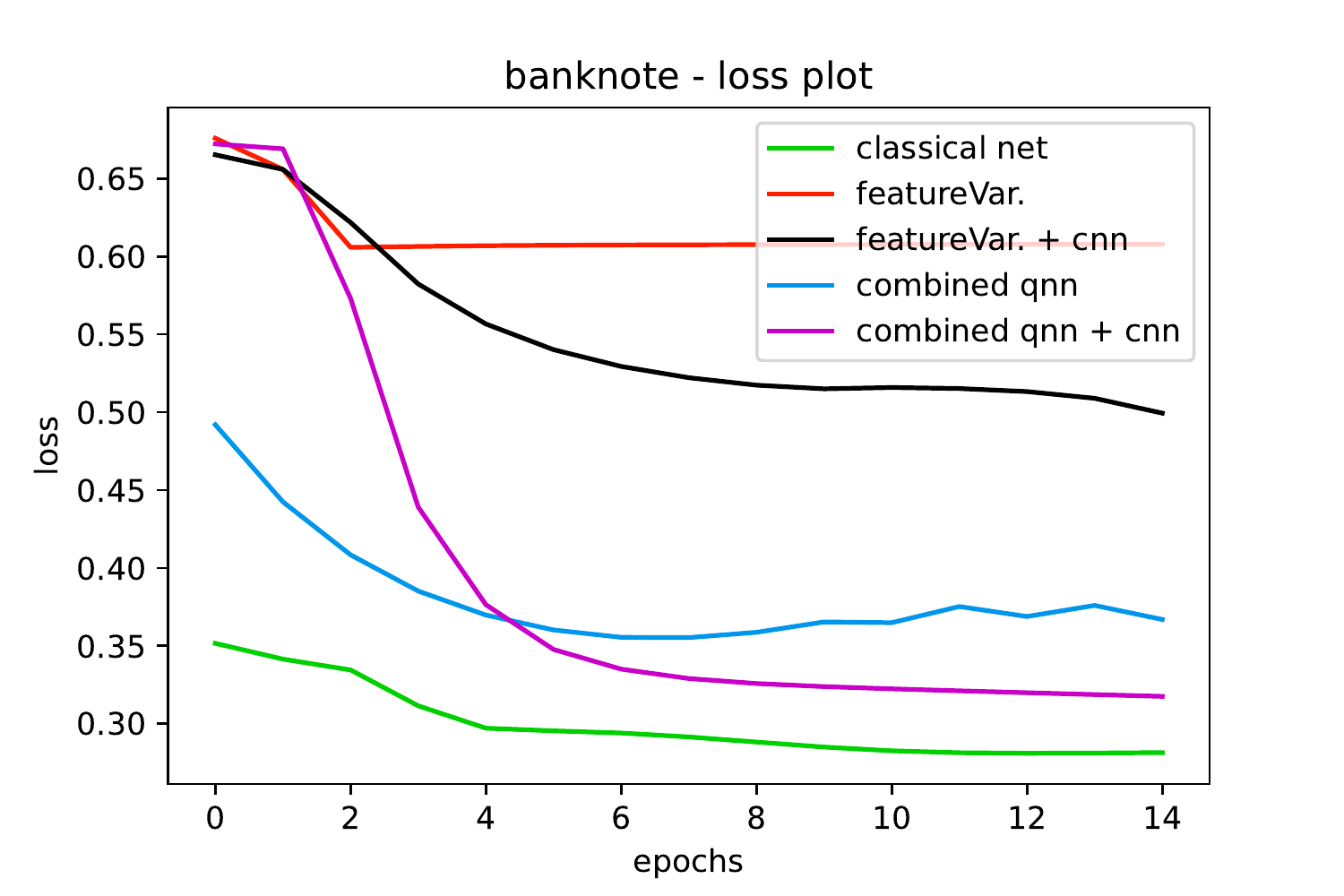}
			\includegraphics[height=3.2in,width=3.2in,angle=0]{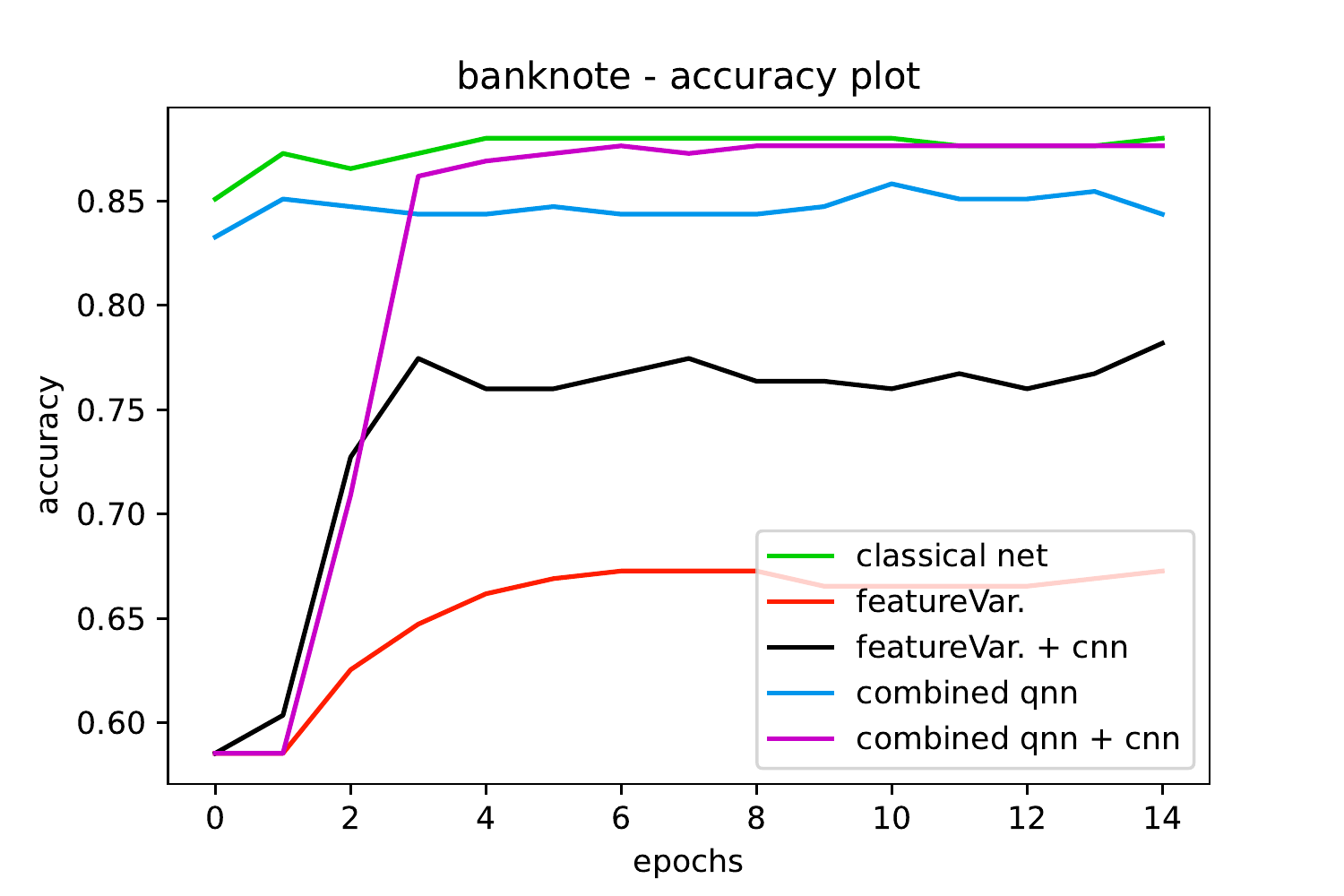}
			\caption{ Plot of the Banknote Fraud dataset results. The binary cross entropy loss (left) and accuracy (right) on the validation set.}
		\end{center}
	\end{figure}

	The \textit{classical net} achieves a loss of 0.300 on the validation set, and an accuracy of 88.4\%. From all models, it scores the best.  The \textit{featureVar.} performs the poorest, with double the loss and an accuracy of only 67.3\%. By adding a classical net to postprocess this qnn (\textit{featureVar. + cnn}), the performance improves but remains below \textit{classical net}. The \textit{combined qnn} and \textit{combined qnn + cnn} achieve results almost on par to the \textit{classical net}, despite a lower parameter count.
	
	\begin{table}[H]
		\begin{tabular}{||c ||c | c | c||} 
			\hline
			model name & \# of classical / quantum parameters  & min loss & max accuracy \\ 
			\hline\hline
			classical net & 15 / 0 & 0.300 & 88.4\%  \\ 
			\hline
			featureVar. & 0 / 4 & 0.606 & 67.3\%  \\
			\hline
			featureVar. + cnn & 3 / 4 & 0.499 & 78.2\%  \\
			\hline
			combined qnn & 0 / 7 & 0.355 & 85.5\%  \\
			\hline
			combined qnn + cnn & 3 / 7 & 0.318 & 87.6\%  \\ [1ex] 
			\hline
		\end{tabular}
		\caption{\label{tab:fraud} Results for the Banknote Fraud dataset. For each model, the number of classical and quantum parameters, the best loss and the best accuracy is shown.   }
	\end{table}

	\section{Discussion}
	Comparing the relative performance of a classical neural network with a quantum one should be done with caution. Firstly, only minimal tuning of hyperparameters (batch size, learning rate, optimizer, ...) was done in the problems studied. Secondly, the number of parameters was not equal across the different models. (A lower parameter count was chosen for the quantum models, to reduce the simulation time. While arguably this penalizes them, on the other hand, the quantum models still are much more costly to run both on NISQs or by simulating them on a classical computer.) Finally, the cases studied here only have a few features, and it is not guaranteed any findings generalize to larger feature spaces. \par
	
	We give a further observation about optimizers: the output of a quantum circuit is inherently stochastic due to the finite amount of runs of the circuit. This is the case even when a quantum computer is fully error corrected. The Adam optimizer used in the simulation wasn't designed with this in mind. To extract the measurement probabilities, we hence use an exact statevector simulation instead of repeatedly running the circuit. While this removes the stochasticity, this would not be possible when running on a real quantum computer. \\

	We use here the classical net as a benchmark for the other models. All three cases showed similar trends:
	
	\begin{enumerate}
		\item The quantum neural networks with separated feature and variational part underperformed in both loss and accuracy.
		\item The combined quantum neural network outperformed the other quantum net designs.
		\item Replacing the fixed post-processing function with a neural net running on a classical computer significantly improved the accuracy and reduced the loss.
		\item The qnn-cnn tandem achieved results on par with the classical net, despite a lower parameter count.
	\end{enumerate}
	
	Some tests (results not shown) were also done by leaving out a hidden layer in the classical net, forming shallower classical neural nets. These model had a lower parameter count, but underperformed and were left out of further analysis.
	
	The key finding is that the new design outperforms the more typical qnn designs and is even on par with deep classical nets. Moreover, the whole `hybrid' (qnn - cnn) setup is more efficient than a classical net, achieving equal results with a much lower amount of parameters.
	
	\section{Conclusion}
	The new architecture proposed within this paper shows promise on a variety of datasets. Merging feature and variational circuit and \textendash \hspace{0.03cm} especially \textendash \hspace{0.03cm} post-processing results with a small classical layer achieves results about equal to deep classical neural networks. It is hypothesized that the tandem of a quantum-classical model can bring out the strengths of both. We suggest blending quantum nets with classical ones as an avenue for further research. \\
	
	\textbf{\textit{Acknowledgments}}
	The author thanks Serge Massar, Marius Petitzon, Matthias Sinnesael, and Sandro Montanari for their helpful comments and their constructive feedback on the manuscript.

	\section* {Code}
	Code to reproduce the datasets and figures can be found in the following GitHub repository: \url{https://github.com/fpetitzon/Study_of_new_designs_in_quantum_neural_networks}

	\section* {Appendix}	
	\label{sec:app}

	\begin{figure}[H]
		\begin{center}
			\includegraphics[angle=0, origin=c, width=1 \textwidth ]{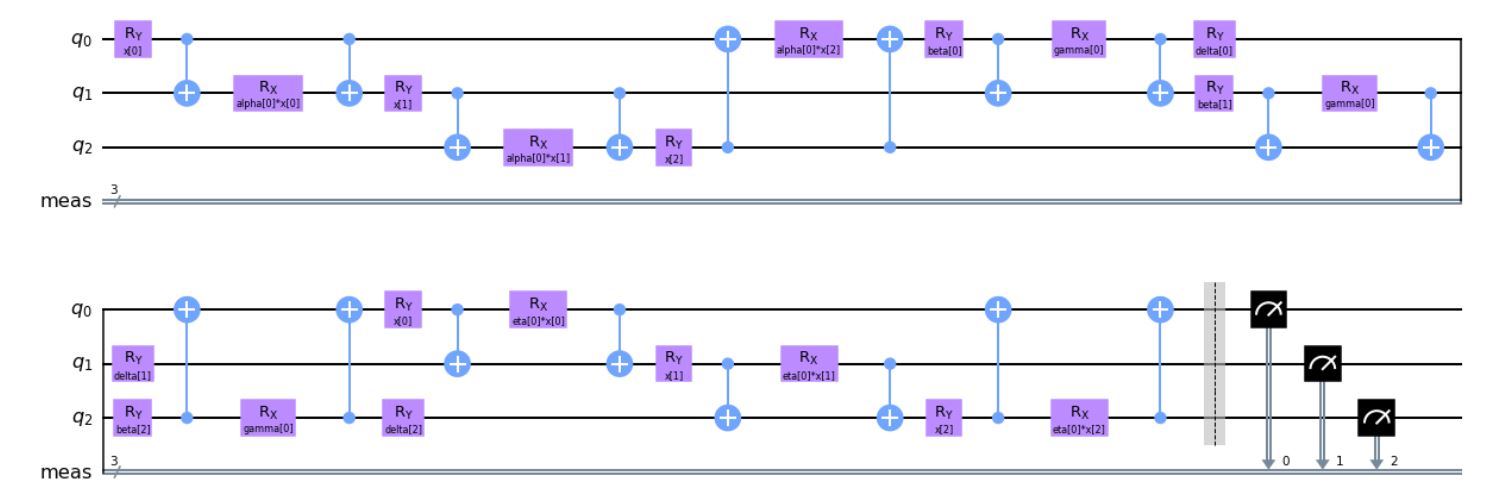}
			\caption{ The combined quantum neural network for the case N = 3. A mix of entanglements and parameterized rotations are executed. X are the input features, and the Greek letters are parameters to be optimized. The number of entanglements scales as $\sim 6N$, and only neighboring connectivity between the qubits is assumed.  }
		\end{center}
	\end{figure}

	\bibliography{refs}

\begin{thebibliography}{24}%
\makeatletter
\providecommand \@ifxundefined [1]{%
 \@ifx{#1\undefined}
}%
\providecommand \@ifnum [1]{%
 \ifnum #1\expandafter \@firstoftwo
 \else \expandafter \@secondoftwo
 \fi
}%
\providecommand \@ifx [1]{%
 \ifx #1\expandafter \@firstoftwo
 \else \expandafter \@secondoftwo
 \fi
}%
\providecommand \natexlab [1]{#1}%
\providecommand \enquote  [1]{``#1''}%
\providecommand \bibnamefont  [1]{#1}%
\providecommand \bibfnamefont [1]{#1}%
\providecommand \citenamefont [1]{#1}%
\providecommand \href@noop [0]{\@secondoftwo}%
\providecommand \href [0]{\begingroup \@sanitize@url \@href}%
\providecommand \@href[1]{\@@startlink{#1}\@@href}%
\providecommand \@@href[1]{\endgroup#1\@@endlink}%
\providecommand \@sanitize@url [0]{\catcode `\\12\catcode `\$12\catcode
  `\&12\catcode `\#12\catcode `\^12\catcode `\_12\catcode `\%12\relax}%
\providecommand \@@startlink[1]{}%
\providecommand \@@endlink[0]{}%
\providecommand \url  [0]{\begingroup\@sanitize@url \@url }%
\providecommand \@url [1]{\endgroup\@href {#1}{\urlprefix }}%
\providecommand \urlprefix  [0]{URL }%
\providecommand \Eprint [0]{\href }%
\providecommand \doibase [0]{http://dx.doi.org/}%
\providecommand \selectlanguage [0]{\@gobble}%
\providecommand \bibinfo  [0]{\@secondoftwo}%
\providecommand \bibfield  [0]{\@secondoftwo}%
\providecommand \translation [1]{[#1]}%
\providecommand \BibitemOpen [0]{}%
\providecommand \bibitemStop [0]{}%
\providecommand \bibitemNoStop [0]{.\EOS\space}%
\providecommand \EOS [0]{\spacefactor3000\relax}%
\providecommand \BibitemShut  [1]{\csname bibitem#1\endcsname}%
\let\auto@bib@innerbib\@empty
\bibitem [{\citenamefont {Dai}\ \emph {et~al.}(2021)\citenamefont {Dai},
  \citenamefont {Liu}, \citenamefont {Le},\ and\ \citenamefont
  {Tan}}]{Dai2021}%
  \BibitemOpen
  \bibfield  {author} {\bibinfo {author} {\bibfnamefont {Z.}~\bibnamefont
  {Dai}}, \bibinfo {author} {\bibfnamefont {H.}~\bibnamefont {Liu}}, \bibinfo
  {author} {\bibfnamefont {Q.~V.}\ \bibnamefont {Le}}, \ and\ \bibinfo {author}
  {\bibfnamefont {M.}~\bibnamefont {Tan}},\ }\bibfield  {title} {\enquote
  {\bibinfo {title} {Coatnet: Marrying convolution and attention for all data
  sizes},}\ }\href@noop {} {\  (\bibinfo {year} {2021})},\ \Eprint
  {http://arxiv.org/abs/2106.04803} {arXiv:2106.04803 [cs.CV]} \BibitemShut
  {NoStop}%
\bibitem [{\citenamefont {Yoo}(2015)}]{Yoo2015}%
  \BibitemOpen
  \bibfield  {author} {\bibinfo {author} {\bibfnamefont {H.-J.}\ \bibnamefont
  {Yoo}},\ }\bibfield  {title} {\enquote {\bibinfo {title} {Deep convolution
  neural networks in computer vision: a review},}\ }\href {\doibase
  10.5573/ieiespc.2015.4.1.035} {\bibfield  {journal} {\bibinfo  {journal}
  {{IEIE} Transactions on Smart Processing and Computing}\ }\textbf {\bibinfo
  {volume} {4}},\ \bibinfo {pages} {35--43} (\bibinfo {year}
  {2015})}\BibitemShut {NoStop}%
\bibitem [{\citenamefont {Deepl}(2022)}]{deepl}%
  \BibitemOpen
  \bibfield  {author} {\bibinfo {author} {\bibnamefont {Deepl}},\ }\href@noop
  {} {}\bibinfo {howpublished} {https://www.deepl.com/en/quality.html}
  (\bibinfo {year} {2022})\BibitemShut {NoStop}%
\bibitem [{\citenamefont {Silver}\ \emph {et~al.}(2016)\citenamefont {Silver},
  \citenamefont {Huang}, \citenamefont {Maddison}, \citenamefont {Guez},
  \citenamefont {Sifre}, \citenamefont {van~den Driessche}, \citenamefont
  {Schrittwieser}, \citenamefont {Antonoglou}, \citenamefont {Panneershelvam},
  \citenamefont {Lanctot}, \citenamefont {Dieleman}, \citenamefont {Grewe},
  \citenamefont {Nham}, \citenamefont {Kalchbrenner}, \citenamefont
  {Sutskever}, \citenamefont {Lillicrap}, \citenamefont {Leach}, \citenamefont
  {Kavukcuoglu}, \citenamefont {Graepel},\ and\ \citenamefont
  {Hassabis}}]{Silver2016}%
  \BibitemOpen
  \bibfield  {author} {\bibinfo {author} {\bibfnamefont {D.}~\bibnamefont
  {Silver}}, \bibinfo {author} {\bibfnamefont {A.}~\bibnamefont {Huang}},
  \bibinfo {author} {\bibfnamefont {C.~J.}\ \bibnamefont {Maddison}}, \bibinfo
  {author} {\bibfnamefont {A.}~\bibnamefont {Guez}}, \bibinfo {author}
  {\bibfnamefont {L.}~\bibnamefont {Sifre}}, \bibinfo {author} {\bibfnamefont
  {G.}~\bibnamefont {van~den Driessche}}, \bibinfo {author} {\bibfnamefont
  {J.}~\bibnamefont {Schrittwieser}}, \bibinfo {author} {\bibfnamefont
  {I.}~\bibnamefont {Antonoglou}}, \bibinfo {author} {\bibfnamefont
  {V.}~\bibnamefont {Panneershelvam}}, \bibinfo {author} {\bibfnamefont
  {M.}~\bibnamefont {Lanctot}}, \bibinfo {author} {\bibfnamefont
  {S.}~\bibnamefont {Dieleman}}, \bibinfo {author} {\bibfnamefont
  {D.}~\bibnamefont {Grewe}}, \bibinfo {author} {\bibfnamefont
  {J.}~\bibnamefont {Nham}}, \bibinfo {author} {\bibfnamefont {N.}~\bibnamefont
  {Kalchbrenner}}, \bibinfo {author} {\bibfnamefont {I.}~\bibnamefont
  {Sutskever}}, \bibinfo {author} {\bibfnamefont {T.}~\bibnamefont
  {Lillicrap}}, \bibinfo {author} {\bibfnamefont {M.}~\bibnamefont {Leach}},
  \bibinfo {author} {\bibfnamefont {K.}~\bibnamefont {Kavukcuoglu}}, \bibinfo
  {author} {\bibfnamefont {T.}~\bibnamefont {Graepel}}, \ and\ \bibinfo
  {author} {\bibfnamefont {D.}~\bibnamefont {Hassabis}},\ }\bibfield  {title}
  {\enquote {\bibinfo {title} {Mastering the game of go with deep neural
  networks and tree search},}\ }\href {\doibase 10.1038/nature16961} {\bibfield
   {journal} {\bibinfo  {journal} {Nature}\ }\textbf {\bibinfo {volume}
  {529}},\ \bibinfo {pages} {484--489} (\bibinfo {year} {2016})}\BibitemShut
  {NoStop}%
\bibitem [{\citenamefont {Jumper}\ \emph {et~al.}(2021)\citenamefont {Jumper},
  \citenamefont {Evans}, \citenamefont {Pritzel}, \citenamefont {Green},
  \citenamefont {Figurnov}, \citenamefont {Ronneberger}, \citenamefont
  {Tunyasuvunakool}, \citenamefont {Bates}, \citenamefont {{\v{Z}}{\'{\i}}dek},
  \citenamefont {Potapenko}, \citenamefont {Bridgland}, \citenamefont {Meyer},
  \citenamefont {Kohl}, \citenamefont {Ballard}, \citenamefont {Cowie},
  \citenamefont {Romera-Paredes}, \citenamefont {Nikolov}, \citenamefont
  {Jain}, \citenamefont {Adler}, \citenamefont {Back}, \citenamefont
  {Petersen}, \citenamefont {Reiman}, \citenamefont {Clancy}, \citenamefont
  {Zielinski}, \citenamefont {Steinegger}, \citenamefont {Pacholska},
  \citenamefont {Berghammer}, \citenamefont {Bodenstein}, \citenamefont
  {Silver}, \citenamefont {Vinyals}, \citenamefont {Senior}, \citenamefont
  {Kavukcuoglu}, \citenamefont {Kohli},\ and\ \citenamefont
  {Hassabis}}]{Jumper2021}%
  \BibitemOpen
  \bibfield  {author} {\bibinfo {author} {\bibfnamefont {J.}~\bibnamefont
  {Jumper}}, \bibinfo {author} {\bibfnamefont {R.}~\bibnamefont {Evans}},
  \bibinfo {author} {\bibfnamefont {A.}~\bibnamefont {Pritzel}}, \bibinfo
  {author} {\bibfnamefont {T.}~\bibnamefont {Green}}, \bibinfo {author}
  {\bibfnamefont {M.}~\bibnamefont {Figurnov}}, \bibinfo {author}
  {\bibfnamefont {O.}~\bibnamefont {Ronneberger}}, \bibinfo {author}
  {\bibfnamefont {K.}~\bibnamefont {Tunyasuvunakool}}, \bibinfo {author}
  {\bibfnamefont {R.}~\bibnamefont {Bates}}, \bibinfo {author} {\bibfnamefont
  {A.}~\bibnamefont {{\v{Z}}{\'{\i}}dek}}, \bibinfo {author} {\bibfnamefont
  {A.}~\bibnamefont {Potapenko}}, \bibinfo {author} {\bibfnamefont
  {A.}~\bibnamefont {Bridgland}}, \bibinfo {author} {\bibfnamefont
  {C.}~\bibnamefont {Meyer}}, \bibinfo {author} {\bibfnamefont {S.~A.~A.}\
  \bibnamefont {Kohl}}, \bibinfo {author} {\bibfnamefont {A.~J.}\ \bibnamefont
  {Ballard}}, \bibinfo {author} {\bibfnamefont {A.}~\bibnamefont {Cowie}},
  \bibinfo {author} {\bibfnamefont {B.}~\bibnamefont {Romera-Paredes}},
  \bibinfo {author} {\bibfnamefont {S.}~\bibnamefont {Nikolov}}, \bibinfo
  {author} {\bibfnamefont {R.}~\bibnamefont {Jain}}, \bibinfo {author}
  {\bibfnamefont {J.}~\bibnamefont {Adler}}, \bibinfo {author} {\bibfnamefont
  {T.}~\bibnamefont {Back}}, \bibinfo {author} {\bibfnamefont {S.}~\bibnamefont
  {Petersen}}, \bibinfo {author} {\bibfnamefont {D.}~\bibnamefont {Reiman}},
  \bibinfo {author} {\bibfnamefont {E.}~\bibnamefont {Clancy}}, \bibinfo
  {author} {\bibfnamefont {M.}~\bibnamefont {Zielinski}}, \bibinfo {author}
  {\bibfnamefont {M.}~\bibnamefont {Steinegger}}, \bibinfo {author}
  {\bibfnamefont {M.}~\bibnamefont {Pacholska}}, \bibinfo {author}
  {\bibfnamefont {T.}~\bibnamefont {Berghammer}}, \bibinfo {author}
  {\bibfnamefont {S.}~\bibnamefont {Bodenstein}}, \bibinfo {author}
  {\bibfnamefont {D.}~\bibnamefont {Silver}}, \bibinfo {author} {\bibfnamefont
  {O.}~\bibnamefont {Vinyals}}, \bibinfo {author} {\bibfnamefont {A.~W.}\
  \bibnamefont {Senior}}, \bibinfo {author} {\bibfnamefont {K.}~\bibnamefont
  {Kavukcuoglu}}, \bibinfo {author} {\bibfnamefont {P.}~\bibnamefont {Kohli}},
  \ and\ \bibinfo {author} {\bibfnamefont {D.}~\bibnamefont {Hassabis}},\
  }\bibfield  {title} {\enquote {\bibinfo {title} {Highly accurate protein
  structure prediction with {AlphaFold}},}\ }\href {\doibase
  10.1038/s41586-021-03819-2} {\bibfield  {journal} {\bibinfo  {journal}
  {Nature}\ }\textbf {\bibinfo {volume} {596}},\ \bibinfo {pages} {583--589}
  (\bibinfo {year} {2021})}\BibitemShut {NoStop}%
\bibitem [{\citenamefont {Feynman}(1982)}]{Feynman1982}%
  \BibitemOpen
  \bibfield  {author} {\bibinfo {author} {\bibfnamefont {R.~P.}\ \bibnamefont
  {Feynman}},\ }\bibfield  {title} {\enquote {\bibinfo {title} {Simulating
  physics with computers},}\ }\href {\doibase 10.1007/bf02650179} {\bibfield
  {journal} {\bibinfo  {journal} {International Journal of Theoretical
  Physics}\ }\textbf {\bibinfo {volume} {21}},\ \bibinfo {pages} {467--488}
  (\bibinfo {year} {1982})}\BibitemShut {NoStop}%
\bibitem [{\citenamefont {Shor}(1994)}]{shor96}%
  \BibitemOpen
  \bibfield  {author} {\bibinfo {author} {\bibfnamefont {P.}~\bibnamefont
  {Shor}},\ }\bibfield  {title} {\enquote {\bibinfo {title} {Algorithms for
  quantum computation: discrete logarithms and factoring},}\ }in\ \href
  {\doibase 10.1109/SFCS.1994.365700} {\emph {\bibinfo {booktitle} {Proceedings
  35th Annual Symposium on Foundations of Computer Science}}}\ (\bibinfo {year}
  {1994})\ pp.\ \bibinfo {pages} {124--134}\BibitemShut {NoStop}%
\bibitem [{\citenamefont {Grover}(1996)}]{grover96}%
  \BibitemOpen
  \bibfield  {author} {\bibinfo {author} {\bibfnamefont {L.~K.}\ \bibnamefont
  {Grover}},\ }\href@noop {} {\enquote {\bibinfo {title} {A fast quantum
  mechanical algorithm for database search},}\ } (\bibinfo {year} {1996}),\
  \Eprint {http://arxiv.org/abs/quant-ph/9605043} {arXiv:quant-ph/9605043
  [quant-ph]} \BibitemShut {NoStop}%
\bibitem [{IBM(2022)}]{IBM2022}%
  \BibitemOpen
  \href@noop {} {}\bibinfo {howpublished} {IBM Quantum, Available online:
  https://quantum-computing.ibm.com/} (\bibinfo {year} {2022})\BibitemShut
  {NoStop}%
\bibitem [{\citenamefont {Rigetti}(2022)}]{Rigetti2022}%
  \BibitemOpen
  \bibfield  {author} {\bibinfo {author} {\bibnamefont {Rigetti}},\ }\href@noop
  {} {\enquote {\bibinfo {title} {Think quantum},}\ }\bibinfo {howpublished}
  {Available online: https://www.rigetti.com} (\bibinfo {year}
  {2022})\BibitemShut {NoStop}%
\bibitem [{\citenamefont {IonQ}(2020)}]{IonQ2020}%
  \BibitemOpen
  \bibfield  {author} {\bibinfo {author} {\bibnamefont {IonQ}},\ }\href@noop {}
  {\enquote {\bibinfo {title} {Ionq unveils world’s most powerful quantum
  computer},}\ }\bibinfo {howpublished}
  {https://ionq.com/news/october-01-2020-most-powerful-quantum-computer}
  (\bibinfo {year} {2020})\BibitemShut {NoStop}%
\bibitem [{\citenamefont {Preskill}(2018)}]{Preskill2018}%
  \BibitemOpen
  \bibfield  {author} {\bibinfo {author} {\bibfnamefont {J.}~\bibnamefont
  {Preskill}},\ }\bibfield  {title} {\enquote {\bibinfo {title} {Quantum
  computing in the nisq era and beyond},}\ }\href {\doibase
  10.22331/q-2018-08-06-79} {\bibfield  {journal} {\bibinfo  {journal}
  {Quantum}\ }\textbf {\bibinfo {volume} {2}},\ \bibinfo {pages} {79} (\bibinfo
  {year} {2018})}\BibitemShut {NoStop}%
\bibitem [{\citenamefont {IBM}(2022{\natexlab{a}})}]{IBM_course_2022}%
  \BibitemOpen
  \bibfield  {author} {\bibinfo {author} {\bibnamefont {IBM}},\ }\href@noop {}
  {\enquote {\bibinfo {title} {Qiskit course, supervized learning},}\ }\bibinfo
  {howpublished}
  {https://learn.qiskit.org/course/machine-learning/supervised-learning,
  accessed 15 Feb 2022} (\bibinfo {year} {2022}{\natexlab{a}})\BibitemShut
  {NoStop}%
\bibitem [{\citenamefont {Benedetti}\ \emph {et~al.}(2019)\citenamefont
  {Benedetti}, \citenamefont {Lloyd}, \citenamefont {Sack},\ and\ \citenamefont
  {Fiorentini}}]{Benedetti2019}%
  \BibitemOpen
  \bibfield  {author} {\bibinfo {author} {\bibfnamefont {M.}~\bibnamefont
  {Benedetti}}, \bibinfo {author} {\bibfnamefont {E.}~\bibnamefont {Lloyd}},
  \bibinfo {author} {\bibfnamefont {S.}~\bibnamefont {Sack}}, \ and\ \bibinfo
  {author} {\bibfnamefont {M.}~\bibnamefont {Fiorentini}},\ }\bibfield  {title}
  {\enquote {\bibinfo {title} {Parameterized quantum circuits as machine
  learning models},}\ }\href {\doibase 10.1088/2058-9565/ab4eb5} {\bibfield
  {journal} {\bibinfo  {journal} {Quantum Science and Technology}\ }\textbf
  {\bibinfo {volume} {4}},\ \bibinfo {pages} {043001} (\bibinfo {year}
  {2019})}\BibitemShut {NoStop}%
\bibitem [{\citenamefont {Schuld}\ \emph {et~al.}(2018)\citenamefont {Schuld},
  \citenamefont {Bocharov}, \citenamefont {Svore},\ and\ \citenamefont
  {Wiebe}}]{Schuld2018}%
  \BibitemOpen
  \bibfield  {author} {\bibinfo {author} {\bibfnamefont {M.}~\bibnamefont
  {Schuld}}, \bibinfo {author} {\bibfnamefont {A.}~\bibnamefont {Bocharov}},
  \bibinfo {author} {\bibfnamefont {K.}~\bibnamefont {Svore}}, \ and\ \bibinfo
  {author} {\bibfnamefont {N.}~\bibnamefont {Wiebe}},\ }\bibfield  {title}
  {\enquote {\bibinfo {title} {Circuit-centric quantum classifiers},}\ }\href
  {\doibase 10.1103/PhysRevA.101.032308} {\bibfield  {journal} {\bibinfo
  {journal} {Phys. Rev. A 101, 032308 (2020)}\ } (\bibinfo {year} {2018}),\
  10.1103/PhysRevA.101.032308},\ \Eprint {http://arxiv.org/abs/1804.00633}
  {arXiv:1804.00633 [quant-ph]} \BibitemShut {NoStop}%
\bibitem [{\citenamefont {Lloyd}\ \emph {et~al.}(2020)\citenamefont {Lloyd},
  \citenamefont {Schuld}, \citenamefont {Ijaz}, \citenamefont {Izaac},\ and\
  \citenamefont {Killoran}}]{Lloyd2020}%
  \BibitemOpen
  \bibfield  {author} {\bibinfo {author} {\bibfnamefont {S.}~\bibnamefont
  {Lloyd}}, \bibinfo {author} {\bibfnamefont {M.}~\bibnamefont {Schuld}},
  \bibinfo {author} {\bibfnamefont {A.}~\bibnamefont {Ijaz}}, \bibinfo {author}
  {\bibfnamefont {J.}~\bibnamefont {Izaac}}, \ and\ \bibinfo {author}
  {\bibfnamefont {N.}~\bibnamefont {Killoran}},\ }\bibfield  {title} {\enquote
  {\bibinfo {title} {Quantum embeddings for machine learning},}\ }\href@noop {}
  {\  (\bibinfo {year} {2020})},\ \Eprint {http://arxiv.org/abs/2001.03622}
  {arXiv:2001.03622 [quant-ph]} \BibitemShut {NoStop}%
\bibitem [{\citenamefont {Cerezo}\ \emph {et~al.}(2021)\citenamefont {Cerezo},
  \citenamefont {Arrasmith}, \citenamefont {Babbush}, \citenamefont {Benjamin},
  \citenamefont {Endo}, \citenamefont {Fujii}, \citenamefont {McClean},
  \citenamefont {Mitarai}, \citenamefont {Yuan}, \citenamefont {Cincio},\ and\
  \citenamefont {Coles}}]{Cerezo2021}%
  \BibitemOpen
  \bibfield  {author} {\bibinfo {author} {\bibfnamefont {M.}~\bibnamefont
  {Cerezo}}, \bibinfo {author} {\bibfnamefont {A.}~\bibnamefont {Arrasmith}},
  \bibinfo {author} {\bibfnamefont {R.}~\bibnamefont {Babbush}}, \bibinfo
  {author} {\bibfnamefont {S.~C.}\ \bibnamefont {Benjamin}}, \bibinfo {author}
  {\bibfnamefont {S.}~\bibnamefont {Endo}}, \bibinfo {author} {\bibfnamefont
  {K.}~\bibnamefont {Fujii}}, \bibinfo {author} {\bibfnamefont {J.~R.}\
  \bibnamefont {McClean}}, \bibinfo {author} {\bibfnamefont {K.}~\bibnamefont
  {Mitarai}}, \bibinfo {author} {\bibfnamefont {X.}~\bibnamefont {Yuan}},
  \bibinfo {author} {\bibfnamefont {L.}~\bibnamefont {Cincio}}, \ and\ \bibinfo
  {author} {\bibfnamefont {P.~J.}\ \bibnamefont {Coles}},\ }\bibfield  {title}
  {\enquote {\bibinfo {title} {Variational quantum algorithms},}\ }\href
  {\doibase 10.1038/s42254-021-00348-9} {\bibfield  {journal} {\bibinfo
  {journal} {Nature Reviews Physics}\ }\textbf {\bibinfo {volume} {3}},\
  \bibinfo {pages} {625--644} (\bibinfo {year} {2021})}\BibitemShut {NoStop}%
\bibitem [{\citenamefont {Havl{\'{\i}}{\v{c}}ek}\ \emph
  {et~al.}(2019)\citenamefont {Havl{\'{\i}}{\v{c}}ek}, \citenamefont
  {C{\'{o}}rcoles}, \citenamefont {Temme}, \citenamefont {Harrow},
  \citenamefont {Kandala}, \citenamefont {Chow},\ and\ \citenamefont
  {Gambetta}}]{Havlicek2019}%
  \BibitemOpen
  \bibfield  {author} {\bibinfo {author} {\bibfnamefont {V.}~\bibnamefont
  {Havl{\'{\i}}{\v{c}}ek}}, \bibinfo {author} {\bibfnamefont {A.~D.}\
  \bibnamefont {C{\'{o}}rcoles}}, \bibinfo {author} {\bibfnamefont
  {K.}~\bibnamefont {Temme}}, \bibinfo {author} {\bibfnamefont {A.~W.}\
  \bibnamefont {Harrow}}, \bibinfo {author} {\bibfnamefont {A.}~\bibnamefont
  {Kandala}}, \bibinfo {author} {\bibfnamefont {J.~M.}\ \bibnamefont {Chow}}, \
  and\ \bibinfo {author} {\bibfnamefont {J.~M.}\ \bibnamefont {Gambetta}},\
  }\bibfield  {title} {\enquote {\bibinfo {title} {Supervised learning with
  quantum-enhanced feature spaces},}\ }\href {\doibase
  10.1038/s41586-019-0980-2} {\bibfield  {journal} {\bibinfo  {journal}
  {Nature}\ }\textbf {\bibinfo {volume} {567}},\ \bibinfo {pages} {209--212}
  (\bibinfo {year} {2019})}\BibitemShut {NoStop}%
\bibitem [{\citenamefont {Gelenbe}, \citenamefont {Mao},\ and\ \citenamefont
  {Li}(1999)}]{Gelenbe1999}%
  \BibitemOpen
  \bibfield  {author} {\bibinfo {author} {\bibfnamefont {E.}~\bibnamefont
  {Gelenbe}}, \bibinfo {author} {\bibfnamefont {Z.-H.}\ \bibnamefont {Mao}}, \
  and\ \bibinfo {author} {\bibfnamefont {Y.-D.}\ \bibnamefont {Li}},\
  }\bibfield  {title} {\enquote {\bibinfo {title} {Function approximation with
  spiked random networks},}\ }\href {\doibase 10.1109/72.737488} {\bibfield
  {journal} {\bibinfo  {journal} {{IEEE} Transactions on Neural Networks}\
  }\textbf {\bibinfo {volume} {10}},\ \bibinfo {pages} {3--9} (\bibinfo {year}
  {1999})}\BibitemShut {NoStop}%
\bibitem [{\citenamefont {Broughton}\ \emph {et~al.}(2020)\citenamefont
  {Broughton}, \citenamefont {Verdon}, \citenamefont {McCourt}, \citenamefont
  {Martinez}, \citenamefont {Yoo}, \citenamefont {Isakov}, \citenamefont
  {Massey}, \citenamefont {Halavati}, \citenamefont {Niu}, \citenamefont
  {Zlokapa}, \citenamefont {Peters}, \citenamefont {Lockwood}, \citenamefont
  {Skolik}, \citenamefont {Jerbi}, \citenamefont {Dunjko}, \citenamefont
  {Leib}, \citenamefont {Streif}, \citenamefont {Dollen}, \citenamefont {Chen},
  \citenamefont {Cao}, \citenamefont {Wiersema}, \citenamefont {Huang},
  \citenamefont {McClean}, \citenamefont {Babbush}, \citenamefont {Boixo},
  \citenamefont {Bacon}, \citenamefont {Ho}, \citenamefont {Neven},\ and\
  \citenamefont {Mohseni}}]{Broughton2020}%
  \BibitemOpen
  \bibfield  {author} {\bibinfo {author} {\bibfnamefont {M.}~\bibnamefont
  {Broughton}}, \bibinfo {author} {\bibfnamefont {G.}~\bibnamefont {Verdon}},
  \bibinfo {author} {\bibfnamefont {T.}~\bibnamefont {McCourt}}, \bibinfo
  {author} {\bibfnamefont {A.~J.}\ \bibnamefont {Martinez}}, \bibinfo {author}
  {\bibfnamefont {J.~H.}\ \bibnamefont {Yoo}}, \bibinfo {author} {\bibfnamefont
  {S.~V.}\ \bibnamefont {Isakov}}, \bibinfo {author} {\bibfnamefont
  {P.}~\bibnamefont {Massey}}, \bibinfo {author} {\bibfnamefont
  {R.}~\bibnamefont {Halavati}}, \bibinfo {author} {\bibfnamefont {M.~Y.}\
  \bibnamefont {Niu}}, \bibinfo {author} {\bibfnamefont {A.}~\bibnamefont
  {Zlokapa}}, \bibinfo {author} {\bibfnamefont {E.}~\bibnamefont {Peters}},
  \bibinfo {author} {\bibfnamefont {O.}~\bibnamefont {Lockwood}}, \bibinfo
  {author} {\bibfnamefont {A.}~\bibnamefont {Skolik}}, \bibinfo {author}
  {\bibfnamefont {S.}~\bibnamefont {Jerbi}}, \bibinfo {author} {\bibfnamefont
  {V.}~\bibnamefont {Dunjko}}, \bibinfo {author} {\bibfnamefont
  {M.}~\bibnamefont {Leib}}, \bibinfo {author} {\bibfnamefont {M.}~\bibnamefont
  {Streif}}, \bibinfo {author} {\bibfnamefont {D.~V.}\ \bibnamefont {Dollen}},
  \bibinfo {author} {\bibfnamefont {H.}~\bibnamefont {Chen}}, \bibinfo {author}
  {\bibfnamefont {S.}~\bibnamefont {Cao}}, \bibinfo {author} {\bibfnamefont
  {R.}~\bibnamefont {Wiersema}}, \bibinfo {author} {\bibfnamefont {H.-Y.}\
  \bibnamefont {Huang}}, \bibinfo {author} {\bibfnamefont {J.~R.}\ \bibnamefont
  {McClean}}, \bibinfo {author} {\bibfnamefont {R.}~\bibnamefont {Babbush}},
  \bibinfo {author} {\bibfnamefont {S.}~\bibnamefont {Boixo}}, \bibinfo
  {author} {\bibfnamefont {D.}~\bibnamefont {Bacon}}, \bibinfo {author}
  {\bibfnamefont {A.~K.}\ \bibnamefont {Ho}}, \bibinfo {author} {\bibfnamefont
  {H.}~\bibnamefont {Neven}}, \ and\ \bibinfo {author} {\bibfnamefont
  {M.}~\bibnamefont {Mohseni}},\ }\bibfield  {title} {\enquote {\bibinfo
  {title} {Tensorflow quantum: A software framework for quantum machine
  learning},}\ }\href@noop {} {\  (\bibinfo {year} {2020})},\ \Eprint
  {http://arxiv.org/abs/2003.02989} {arXiv:2003.02989 [quant-ph]} \BibitemShut
  {NoStop}%
\bibitem [{\citenamefont {IBM}(2022{\natexlab{b}})}]{ZzFeatureIbm}%
  \BibitemOpen
  \bibfield  {author} {\bibinfo {author} {\bibnamefont {IBM}},\ }\href@noop {}
  {\enquote {\bibinfo {title} {Qiskit - zzfeaturemap},}\ }\bibinfo
  {howpublished}
  {https://qiskit.org/documentation/stubs/qiskit.circuit.library.ZZFeatureMap.html,
  accessed 15 Feb 2022} (\bibinfo {year} {2022}{\natexlab{b}})\BibitemShut
  {NoStop}%
\bibitem [{\citenamefont {IBM}(2022{\natexlab{c}})}]{RealAmpIbm}%
  \BibitemOpen
  \bibfield  {author} {\bibinfo {author} {\bibnamefont {IBM}},\ }\href@noop {}
  {\enquote {\bibinfo {title} {Qiskit - realamplitudes},}\ }\bibinfo
  {howpublished}
  {https://qiskit.org/documentation/stubs/qiskit.circuit.library.RealAmplitudes.html,
  accessed on 15 Feb 2022} (\bibinfo {year} {2022}{\natexlab{c}})\BibitemShut
  {NoStop}%
\bibitem [{dia(2022)}]{diabetes2022}%
  \BibitemOpen
  \href@noop {} {\enquote {\bibinfo {title} {Pima indians diabetes database},}\
  }\bibinfo {howpublished}
  {https://www.kaggle.com/uciml/pima-indians-diabetes-database, accessed on 15
  Feb 2022} (\bibinfo {year} {2022})\BibitemShut {NoStop}%
\bibitem [{\citenamefont {Dua}\ and\ \citenamefont {Graff}(2017)}]{Dua:2019}%
  \BibitemOpen
  \bibfield  {author} {\bibinfo {author} {\bibfnamefont {D.}~\bibnamefont
  {Dua}}\ and\ \bibinfo {author} {\bibfnamefont {C.}~\bibnamefont {Graff}},\
  }\href {http://archive.ics.uci.edu/ml} {\enquote {\bibinfo {title} {Uci
  machine learning repository},}\ } (\bibinfo {year} {2017})\BibitemShut
  {NoStop}%
\end{thebibliography}%

\end{document}